\documentclass[aps,twocolumn,showpacs,preprintnumbers, floats]{revtex4}
\usepackage{amsfonts}
\usepackage{amsmath}
\usepackage{graphicx}
\usepackage{dcolumn}
\usepackage{bm}

\input{tcilatex}

\begin{document}

\preprint{}
\title{Strain Gradients in Epitaxial Ferroelectrics}
\author{G. Catalan$^{1}$}
\email{g.catalan@chem.rug.nl}
\author{B. Noheda$^{1}$}
\author{J. McAneney$^{2}$}
\author{L. Sinnamon$^{2}$}
\author{J. M. Gregg$^{2}$}
\affiliation{$^{1}$Materials Science Center, University of
Groningen, Groningen 9747AG, Netherlands}
\affiliation{$^{2}$Department of Pure and Applied Physics, Queen's
University Belfast, Belfast BT7 1NN, UK}
\date{November 18, 2004}

\begin{abstract}
X-ray analysis of ferroelectric thin layers of Ba$_{1/2}$Sr$_{1/2}$TiO$%
_{3}$ with different thickness reveals the presence of internal
strain gradients across the film thickness and allows us to
propose a functional form for the internal strain profile. We use
this to calculate the direct influence of strain gradient, through
flexoelectric coupling, on the degradation of the ferroelectric
properties of thin films with decreasing thickness, in excellent
agreement with the observed behaviour. This work highlights the
link between strain relaxation and strain gradients in epitaxial
films, and shows the pressing need to avoid strain gradients in
order to obtain thin ferroelectrics with bulk-like properties.
\end{abstract}

 \maketitle

The interest in ferroelectric thin films is rapidly expanding due
to the recent development of both experimental techniques and
calculation tools that allow to explore the ferroelectric
phenomena at atomic level \cite{Ramesh02, Ahn04}. The
incorporation of realistic mechanical and electrical boundary
conditions in the first-principles formulations is generating new
insight on the mechanisms limiting the ferroelectric response in
thin ferroelectric layers \cite{Junquera03,Kornev04,Bungaro04}.
But while the new evidence suggests that ferroelectricity may
indeed be stable in thin films only a few monolayers thick
\cite{Ramesh02,Ahn04,Junquera03}, the sharp peak in dielectric
constant usually associated with the ferroelectric transition is
systematically depressed in thin films. This obviously limits the
technological impact that would arise from the ability to maintain
ferroelectricity and large dielectric constants down to the
nanoscale in real devices.

Strain caused by lattice mismatch with the substrate is an
important factor affecting the properties of thin films. Strain
can modify the phase diagram of epitaxial ferroelectrics
\cite{Dieguez04, Pertsev98}, change the order of the
transition\cite{Pertsev98, Basceri97}, and shift transition temperatures \cite%
{Pertsev98, Sinnamon02}. However, strain alone does not generally
account for the observed smearing of the dielectric peak, as a
sharp anomaly is still expected at the strain-modified transition
temperature. Gradient terms (of strain, composition, defects, etc)
have recently been proposed to account for the reduced dielectric
constant \cite{Catalan04, Bratkovsky04}. However, no experimental
studies have provided quantitative insight in the gradient terms.
The aim of the present work is to detect and measure strain
gradients in a set of lattice-mismatched epitaxial thin films, and
to correlate the measured gradients with the measured dielectric
properties. The tools used in this work can be applied to any
material where gradients, not just of strain but also of
impurities or vacancies, are expected play a role. Showing the
link between strain relaxation and strain gradients has therefore
wider implications beyond ferroelectricity and is an important
result for general thin film epitaxy.

The films studied in this work are Ba$_{0.5}$Sr$_{0.5}$TiO$_{3}$
(BST) dielectric layers with SrRuO$_{3}$ (SRO) bottom electrodes,
grown by pulsed laser deposition. Epitaxy has been verified by
cross-sectional high-resolution transmission electron microscopy
(TEM). Details of the growth and TEM characterisation are
published elsewhere \cite{Sinnamon02}. In the present work the
crystallographic analysis has been performed using a Philips
X'pert MRD diffractometer with CuK$\alpha _{1}$radiation ($\lambda
$=1.540 \AA ).

\begin{figure}
\includegraphics[scale=0.75]{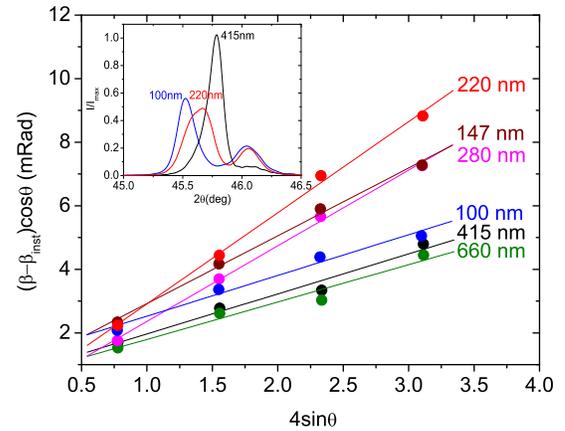}
\caption{(Color) Williamson-Hall plots as a function of
film-thickness. Inset: Diffraction peaks show that broadening is
maximum at intermediate thickness. The peaks at 46.1 deg.
correspond to the electrode (SRO)}
\end{figure}

The lattice parameters are extracted from the position of the
perovskite pseudocubic (002) diffraction peak (see inset of Figure
1). This allows the
calculation of the average out-of-plane strain in each film, given by $\bar{\epsilon}(t)=%
\frac{\bar{c}(t)-c_{0}}{c_{0}}$, where $\overline{c}$\ is the
average out-of-plane lattice parameter, c$_{0}$ is the reference
value, i.e., the bulk lattice parameter (c$_{BST}$= 3.95 \AA ),
and \textit{t} is the film thickness. The average strain for each
film is shown in the inset of Figure 2. There is an out-of-plane
expansion for the thinnest films that decreases with thickness, as
observed before \cite{Sinnamon02, Canedy00}. This is consistent
with relaxation of the in-plane compression induced by the smaller
lattice parameter of the bottom electrode (c$_{SRO}$ =3.93\AA ).

Similar to what is known for semiconductor and metallic epitaxial
layers, perovskite oxides are known to relieve strain as film
thickness is increased. The strain-relieving mechanism is thought
to be mainly the formation of misfit dislocations. As thickness
increases, the accumulation of elastic strain energy overcomes the
barrier for the formation of misfit dislocations, which ease the
strain. It is generally implied that this strain relaxation is
homogeneous across the film,
and thus the strain state is only a function of the film's total thickness: $%
\epsilon $=$\epsilon $(\textit{t}).

However, around a dislocation the lattice is locally distorted.
The accumulation of misfit dislocations at the film-substrate
interface means that the film's lattice parameters are more
distorted near the clamped surface than at the free one, leading
to stress/strain distributions \cite{Canedy00, Chu04, Alpay04,
Nicola04}. Furthermore, strain may not be relaxed solely by
dislocations, as other inhomogeneous mechanisms (such as vertical
segregation of different-sized cations) have been observed
\cite{Maurice03}. Thus, rather than a quantity dependent only on
the thickness \textit{t}, strain should be described as an
internal profile dependent also on the distance to the
film-substrate interface, \textit{z}: $\epsilon $=$\epsilon
$(\textit{z,t}).

The homogeneous vs inhomogeneous scenarios of strain relaxation
are not only different from a structural point of view, but have
consequences for the functional properties. Inhomogeneous strain
fields around dislocations \cite{Chu04, Alpay04} and impurities
\cite{Balzar04} affect the polarisation and critical temperatures
of ferroelectric thin films. Crucially, also, inhomogeneous strain
is necessarily associated with local strain gradients, which
couple to the polarisation via the flexoelectric effect
\cite{Kogan64, Ma01, Tagantsev86}. Measuring the vertical strain
gradient is therefore essential to correctly describe the
functional properties of ferroelectric thin films.

In order to calculate the strain gradients, x-ray diffraction peak
broadening has been analysed as a function of film thickness.
There are at least two contributions to peak broadening: one due
to the finite thickness of the sample, and another due to the
inhomogeneous strain. The two have different angular dependence,
and can therefore be separated by looking at peak width for
different reflections and fitting the results using the
Williamson-Hall relation \cite{WH53}:
\begin{equation}
\beta \cos \theta =K\frac{\lambda }{D}+4\epsilon _{i}\sin (\theta ) .
\label{WH}
\end{equation}%
where \textit{D} is the coherent length perpendicular to the
film's surface (roughly proportional to the film's thickness),
$\lambda $ is the X-ray wavelength ($\lambda $=1.54\AA\ in our
case), $\theta $ is the diffraction angle, $\beta $ is the peak
integral breadth (close to the full width half maximum) minus the
instrumental broadening, and \textit{K} is an empirical constant
close to 1.

Linear fits of $\beta$cos$\theta$ vs sin$\theta$ yield the
coherent length \textit{D} and inhomogeneous
strain $\epsilon _{i}$ for each film. We have performed such fits for the
(00h) (h=1:4) reflections, finding the linear dependence excellent for all our samples (r$^{2}$$>$%
0.9)(Figure 1). It is nevertheless worth mentioning that the
linear W-H plots are only one of existing strategies to separate
size and strain broadening. Quantitative results for the
inhomogeneous strain may
therefore vary depending on the approach used \cite%
{Berkum96}.
\begin{figure}
\includegraphics[scale=0.85]{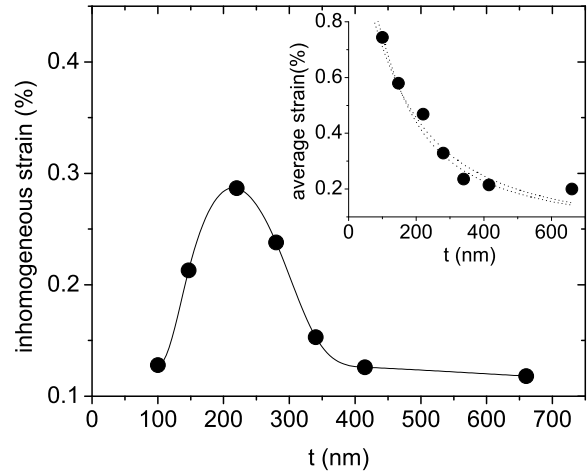}
\caption{Inhomogeneous strain as a function of film thickness. The
solid line is a visual guide. Inset: average out-of-plane strain.
Dotted lines are least-square fits using the averages of Eqs.
\protect\ref{exponential} and \protect\ref{hiperbolic}.}
\end{figure}

Figure 2 shows the inhomogeneous strain as a function of film
thickness. The existence of a maximum in $\epsilon _{i}$ confirms
the presence of a monotonically decreasing internal strain profile
as a function of \textit{z} in the films. For very thin films the
reduced thickness means a small dispersion in lattice parameters;
conversely, for very thick films there may be a large difference
between the lattice parameters at the clamped and free interfaces,
but the volume fraction of totally relaxed material is large and
dominates the diffraction, so that again the variance is small. In
between, there is an intermediate thickness where there is a large
dispersion in lattice parameters, with their thickness fractions
being similar. At this point the inhomogeneous strain is maximum.

Establishing the actual mechanism of strain relaxation is beyond
the scope of this paper, but it is worth mentioning that the
observed behaviour is consistent with predictions for
dislocation-based relaxation: the diffraction peak width in this case is expected to be proportional to $\sqrt{%
\rho /t}$, where $\rho $ is the linear dislocation density, which
grows rapidly around a critical thickness and then saturates
\cite{Kaganer97}.

Extracting quantitative values for the internal strain
profile from this analysis requires solving the integral equations
for the average ($\overline{\epsilon }$) and inhomogeneous
($\epsilon _{i}$) strain:
\begin{eqnarray}
\bar{\epsilon}(t)=\frac{1}{t}\int_{0}^{t}\epsilon (z,t)dz,  \label{average}
\\
\epsilon _{i}^{2}=\frac{1}{t}\int_{0}^{t}\left[ \epsilon ^{2}(z,t)-\bar{%
\epsilon}^{2}(t)\right] dz=\overline{\epsilon ^{2}}-\bar{\epsilon}^{2}
\label{inhomogeneous}
\end{eqnarray}%
where $\overline{\epsilon }$, $\epsilon _{i}$ are extracted from
peak position and width, respectively, and $\epsilon
(z,t)=\frac{c(z,t)-c_{0}}{c_{0}}$ is the internal strain profile.

The easiest way to resolve these equations is the inverse approach: assume a
certain shape for the internal strain, solve the integrals (\ref{average}%
), (\ref{inhomogeneous}), and modify the functional parameters to
achieve a good match with the experimental results. This method
relies heavily on the correct choice of functional dependence for
$\epsilon $(\textit{z}, \textit{t}). As such, the results of the
quantitative analysis should be treated only as approximations.

A general model for the strain profile, independent of the actual
relaxation mechanism, reflects that strain relaxation should be
proportional to the strain itself, which yields an exponential
dependence on \textit{z} \cite{Kim99}:
\begin{equation}
\frac{\partial \epsilon }{\partial z}=-\frac{\epsilon }{\delta }%
\Longrightarrow \epsilon (z)=\epsilon _{0}e^{-\frac{z}{\delta}}
\label{exponential}
\end{equation}%
where $\epsilon _{0}$ is the strain at the film-substrate interface, and $%
\delta $ is a measure of the penetration depth of the strain. If
dislocations are considered as the main relaxation mechanism, a
recent strain-gradient theory \cite{Nicola04} predicts the
vertical profile in the layers to be given by
\begin{equation}
\epsilon (z,t)=\epsilon _{0}\left[ \cosh \frac{z}{\delta }-\tanh \frac{t}{%
\delta }\sinh \frac{z}{\delta }\right]  \label{hiperbolic}
\end{equation}

It is worth noticing that Eq. (\ref{exponential}) is a limiting
case of (\ref{hiperbolic}) when the film thickness is larger than
the strain penetration depth (t$\gg \delta $).

Either of these expressions can be integrated to yield $%
\overline{\epsilon }(t)$. The least-squares fits to the
experimental results using both are shown as dashed lines in the
inset of figure 1. The value of the fitting parameters is $\epsilon _{0}=0.013\pm0.001$ and $%
\delta=60\pm12$ nm, and $\epsilon _{0}$=0.010$\pm$0.001 and $%
\delta$=85$\pm$13 nm for the fits with the averages of
(\ref{exponential}) and (\ref{hiperbolic}) respectively.

The calculated curves for $\epsilon _{i}$, using the above
parameters also reproduce the experimental results for the maximum
value and associated thickness. However, beyond the maximum, the
predicted relaxation of $\epsilon _{i}$ using these equations is
slower than experimentally measured. This discrepancy may be
explained by the presence of more than one relaxation mechanism,
each with different penetration length $\delta $. Furthermore,
$\epsilon _{0}$ is a function of \textit{t}, since the increase in
dislocation density for thicker films affects the strain at the
film-substrate interface. Thus, both parameters should, in
principle, be considered as thickness-dependent: $\epsilon
_{0}$(\textit{t}), $\delta$(\textit{t}). In order to calculate the
thickness dependence of $\epsilon _{0}$(\textit{t}) and $\delta
$(\textit{t}) we note that there are two parameters and two
equations to describe $\overline{\epsilon}$ and $\epsilon _{i}$,
so it is possible to calculate $\epsilon _{0}$ and $\delta $ for
each film separately. We have done this for the exponential strain
profile (\ref{exponential}). Combining
the Eqs. (\ref{average}) and (\ref{inhomogeneous}) we can eliminate $%
\epsilon _{0}$:
\begin{equation}
\frac{t}{2\delta}\frac{\bar{\epsilon}^{2}(t)}{\left( \epsilon _{i}^{2}+\bar{%
\epsilon}^{2}\right) }=\tanh \left( \frac{t}{2\delta}\right)
\label{parameters}
\end{equation}

This is solved for each film in order to find $\delta
$(\textit{t}), which is then used to calculate $\epsilon _{0}(t)$.

Once the internal strain profile $\epsilon $(z,t) is known, the
strain gradient contribution to the functional properties can be
calculated using an elastodielectric free energy expansion
incorporating the flexoelectric contribution:
\begin{widetext}
\begin{equation}
G=\int_{0}^{t}\left[ \frac{1}{2}aP^{2}+\frac{1}{4}bP^{4}-\frac{1}{2}%
(s_{11}+s_{12})\sigma ^{2}-Q_{13}\sigma P^{2}-\gamma
P\frac{\partial \sigma }{\partial z}-\eta \sigma \frac{\partial
P}{\partial z}+\frac{1}{2}C\left(
\frac{\partial P}{\partial z}\right)^{2}+D\left(\frac{\partial \sigma }{%
\partial z}\right) ^{2}\right] dz\label{Landau}
\end{equation}%
\end{widetext}where \textit{P} is the out-of-plane polarisation; \textit{s}$_{ij}$
the elastic compliances; $\sigma $ the in-plane stress (related to
the measured out-of-plane strain by the Young's modulus and
Poisson's ratio: $\sigma=\epsilon$Y(-2$\nu$)); \textit{Q}$_{13}$
is the transverse electrostrictive coefficient, C and D are the
constants related to the energy contributions from polarisation
and stress gradient, and $\gamma $ and $\nu $ are, respectively,
the direct and converse flexoelectric coefficients. \textit{P} is
calculated by minimising the thermodynamic potential, while the
second derivative of \textit{G} with respect to \textit{P} yields
the inverse permittivity. This is averaged over the thickness of
the film to yield the effective value. The values of the
coefficients used in this expansion are the same as in
\cite{Catalan04}.
\begin{figure}
\includegraphics[scale=1.0]{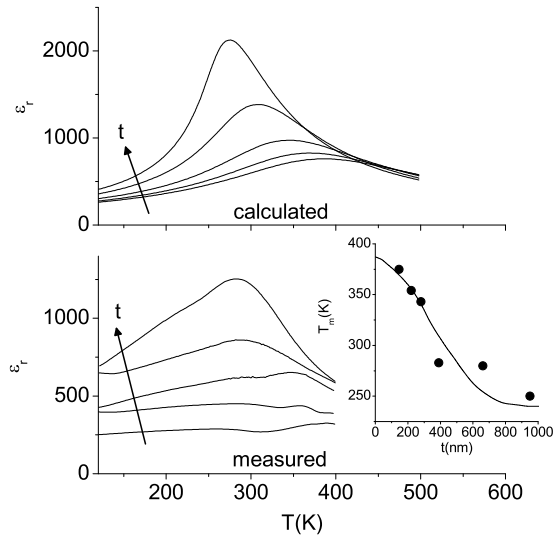}
\caption{Calculated and measured relative dielectric constant as a
function of temperature for films of thickness 660 nm, 340nm,
280nm, 220nm and 145nm. Inset: Temperature of maximum
permittivity, experimental (dots) and calculated (solid line)}
\end{figure}

The relative dielectric constants calculated using the strain
gradient extracted from our crystallographic analysis are shown in
Fig.(2), along with experimentally measured for the same set of
films. The predicted and measured temperatures of maximum
permittivity (\textit{T}$_{m}$) are shown in the inset.

Clearly, the decrease in dielectric constant and upward shift of
\textit{T}$_{m}$ are well reproduced. Quantitatively, the
prediction for \textit{T}$_{m}$ as a function of thickness is
remarkably good, while the calculated dielectric constant is
larger than experimentally measured. This was expected, as our
model does not take into consideration any other
permittivity-depressing factors. The results show that the
contribution to the depression in permittivity with decreasing
thickness (size effect) from flexoelectricity alone is enormous. This is particularly valid
when compared with the huge permittivities recently
measured in gradient-free ferroelectric films \cite{Saad04}.

It is worth noting that the dielectric constant is lowest for the
thinnest films in spite of the relatively small value of $\epsilon
_{i}$. This is a natural consequence of the fact that the size
effect is \textit{not} caused by the inhomogeneous strain itself,
but by the strain gradient, which is largest for the thinnest
films. We emphasise also that while compressive in-plane strain
can indeed be used to stabilise the ferroelectric state, this may
come at the expense of reducing the permittivity if strain
gradients are not avoided. Also, the procedure outlined here could
be used to estimate vertical oxygen vacancy distributions, or
gradients due to impurity concentration. Finally, these methods
open the scope for studying the effect of strain gradients on
other functional materials.

In summary, X-ray analysis of peak broadening as a function of
thickness shows that relaxation of strain in epitaxial films is
associated with the appearance of internal strain gradients. The
dielectric constants calculated using these strain gradients are
close to experimentally measured, clearly showing the fundamental
role plaid by flexoelectric coupling in decreasing the dielectric
constant. This work shows the urgent need to avoid strain
gradients in order to prevent degradation of the ferroelectric
response in thin films.

Useful discussions with T. Hibma, E. Van der Giessen, T. Palstra
and D. Boer are gratefully acknowledged.

\end{document}